\documentclass[12pt,twocolumn]{aastex631}

\usepackage[english]{babel}
\usepackage{amsmath,amssymb,amsfonts, bm,bbm,slashed, subdepth}
\usepackage{mathalfa}
\usepackage{graphicx}
\usepackage{hyperref}
\usepackage{enumerate}
\usepackage{epsfig, subfigure}
\usepackage{multirow}
\usepackage{booktabs}

\usepackage{braket}

\allowdisplaybreaks

\DeclareMathAlphabet\mathbfcal{OMS}{cmsy}{b}{n}

\let\oldhat\hat

\renewcommand{\hat}[1]{\bm\oldhat{\mathbf{#1}}}

\newcommand{\gff}{g_\text{ff}}

\newcommand{\ie}{\textit{i.e.}}
\newcommand{\eg}{\textit{e.g.}}

\begin{document}

\title{Non-relativistic electron-ion bremsstrahlung: an approximate formula for all parameters}

\author[0000-0002-9165-4813]{Josef Pradler}
\affiliation{Institute of High Energy Physics, Austrian Academy of Sciences, Nikolsdorfergasse 18, 1050 Vienna, Austria}
\email{josef.pradler@oeaw.ac.at}
\author[0000-0002-7059-2094]{Lukas Semmelrock}
\affiliation{Institute of High Energy Physics, Austrian Academy of Sciences, Nikolsdorfergasse 18, 1050 Vienna, Austria}
\affiliation{AIT Austrian Institute of Technology, Giefinggasse 4, 1210 Vienna, Austria}
\email{lukas.semmelrock@oeaw.ac.at}

\begin{abstract}
The evaluation of the electron-ion bremsstrahlung cross section---exact to all orders in the Coulomb potential---is computationally expensive 
due to the appearance of hypergeometric functions.
Therefore tabulations are widely used. 
Here, we provide an approximate formula for the non-relativistic dipole process  valid for all applicable relative velocities and photon energies. Its validity spans from the Born- to the classical regime and from soft-photon emission to the kinematic endpoint. The error remains below 3\% (and widely below 1\%) except at an isolated region of hard-photon emission at the quantum-to-classical crossover. We use the formula to obtain the 
thermally averaged emission spectrum and cooling function in a Maxwellian plasma and demonstrate that they are accurate to better than~2\%.
\vspace{30pt}
\end{abstract}

\section{Introduction}

The problem of electron-ion bremsstrahlung has a long history and is of ample importance in astrophysics among other branches of physics.
Early classical and quantum mechanical calculations date back almost a century~\citep{Kramers:1923,Oppenheimer1929, Sugiura1929,Sommerfeld1931,Bethe:1934za}. 
The full relativistic process was treated by \cite{1969PhRv..183...90E} who obtained a triple differential cross section in scattering angle, photon emission angle and energy. Although this result is considered state-of-the-art in calculating free-free transitions of electrons and ions \citep{Chluba:2019ser}, it is obtained in an expansion of $(Z\alpha)$, where $Z$ is the ion charge and $\alpha$ is the fine-structure constant,
on the account that a closed form continuum solution of the Dirac equation for a pure Coulomb field remains unknown~\citep{Bethe:1954zz}. However, when relativistic corrections are small, a non-relativistic single differential cross section in photon energy, exact in the mutual Coulomb interaction of the colliding electron-ion pair, exists~\citep{Sommerfeld:1935ab}.

The latter result involves the evaulation of hypergeometric functions over a wide range of parameters and arguments, 
which can become computationally prohibitive in certain kinematic regimes. 
For this reason, dedicated tabulations of the associated Gaunt factor have been presented~\citep{1961ApJS....6..167K,vanHoof:2014bha,vanHoof:2015jma} that enter, \eg, spectral synthesis codes such as~\texttt{CLOUDY}~\citep{2013RMxAA..49..137F,2017RMxAA..53..385F}.
However, it is desirable for many applications to have simple analytic formul\ae\ at hand that allow for fast approximate calculations, even if they cannot compete in terms of computational speed and accuracy with high-precision table-lookup and interpolation.  Indeed, a sizable number of approximations with varying, but always confined regions of validity exist for the non-relativistic regime; for an overview, see~\cite{1962RvMP...34..507B}.

The  dimensionless numbers that characterize the Coulomb bremsstrahlung process are the Sommerfeld parameters $\nu_{i,f}$ and the 
emitted energy fraction~$x$,
\begin{align}
    \nu_{i,f} \equiv \frac{\alpha Z }{ v_{i,f}},\quad
    x\equiv \frac{2 \omega}{m_e v_{i}^2}, 
\end{align}
where $v_{i,f}$ is the initial (final) non-relativistic electron velocity, $m_e$ is the electron mass and $\omega$ is the energy of the emitted photon; note that $\nu_f  = \nu_i/\sqrt{1-x}$.
The known approximations apply to special values of $x$ and $\nu_{i,f}$.
The Born expressions for $\nu_{i,f} \ll 1$~\citep{Oppenheimer1929, Sugiura1929} were extended 
 to $\nu_i\lesssim 1$ and arbitrary $\nu_f$ by~\cite{Elwert:1939km}. In the  classical regime $\hbar\to 0$, the Sommerfeld parameters become large, $\nu_{i,f} \gg 1$, and the emitted photon energy becomes small, $x\ll 1$, while the product $x\nu_i$ remains independent of $\hbar$.
 In this region of parameter space, there then exist two classical limits, one for $x\nu_i\ll1$~\citep{landau1975classical} and one for $x\nu_i\gg1$~\citep{Kramers:1923}. Only most recently, a soft-photon formula ($x\ll1$) was found that works for arbitrary $\nu_i$~\citep{Weinberg:2019mai}, hence connecting the Born and classical regimes.

All of the above asymptotic expressions and approximations work well in their regions of validity, but have large errors outside, and are therefore only of limited use when calculating integrated quantities like emission spectra and cooling rates. It is the purpose of this paper, to present an approximate formula that works for arbitrary $\nu_i$ and all allowed values of $x$, hence spanning the entire non-relativistic parameter space. Although the formula cannot be derived from first principles, it lends itself to easy, yet reasonably accurate determinations of 
the energy-differential photon emission cross section, of its thermally averaged counterpart as well as of the cooling function.
For example, it may be used to  gain insights into scaling relations or to verify the implementation of tabulated rates.
The formula is hence a first of its kind for the important  free-free dipole process, and we test its validity by comparing against the exact results. A similar program, based on a soft-photon formula obtained in~\cite{Pradler:2020znn}, was recently carried out for non-relativistic quadrupole bremsstrahlung in~\cite{Pradler:2021ppc}.

The paper is organized as follows: 
in Sec.~\ref{sec:ff_gaunt} we introduce the Gaunt factor for electron-ion bremsstrahlung and compare the exact literature result to our approximate formula.
In Sec.~\ref{sec:therm_gaunt} we provide thermal averages in a Maxwellian plasma, obtain the production spectrum and cooling function.
Conclusions are offered in Sec.~\ref{sec:conclusions}.
Throughout the paper we use natural units $\hbar = c = k_B =1$.

\section{Free-Free Gaunt Factor}
\label{sec:ff_gaunt}

\begin{figure*}[tb]
        \includegraphics[height=0.83\columnwidth]{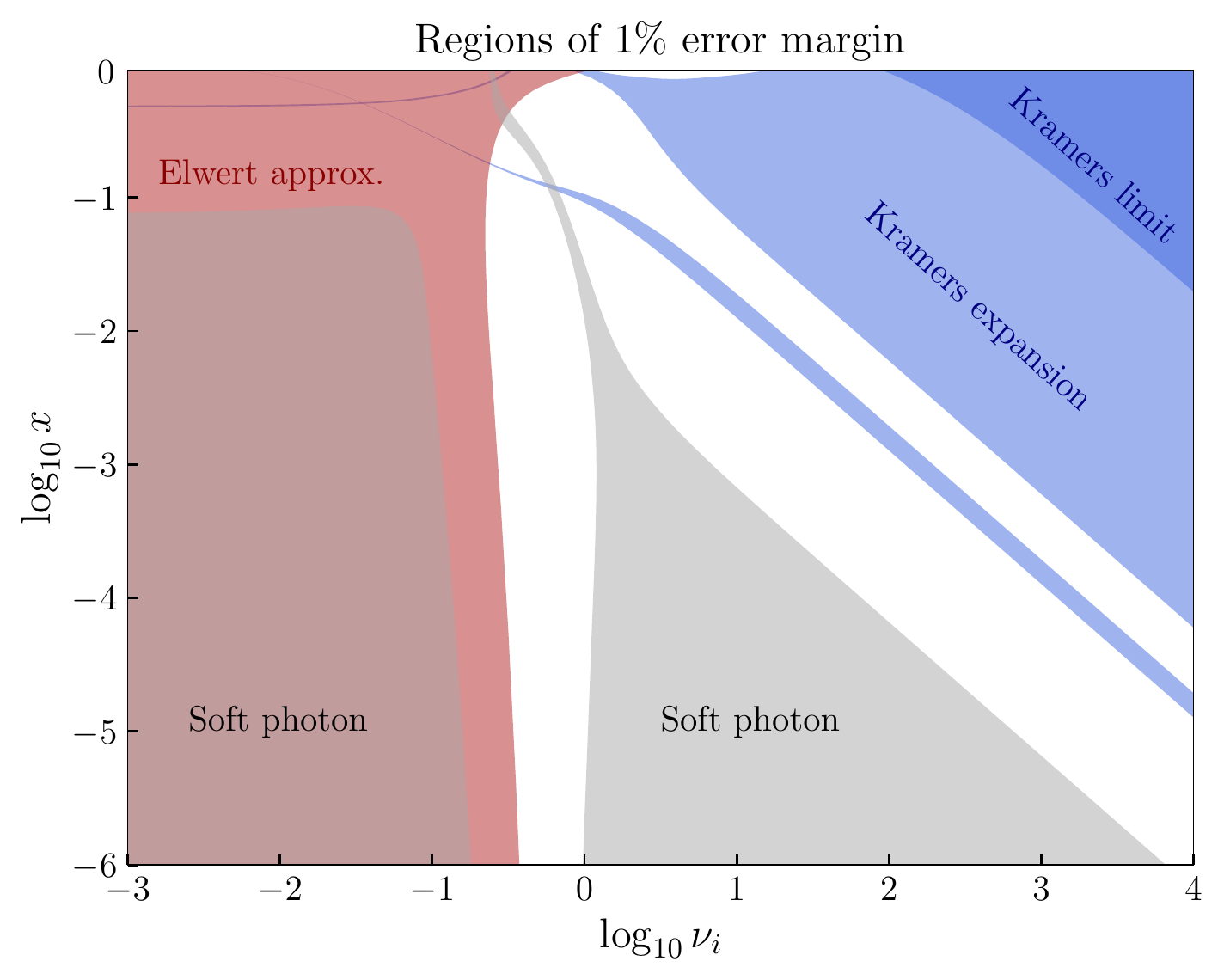}\hfill
 \includegraphics[height=0.83\columnwidth]{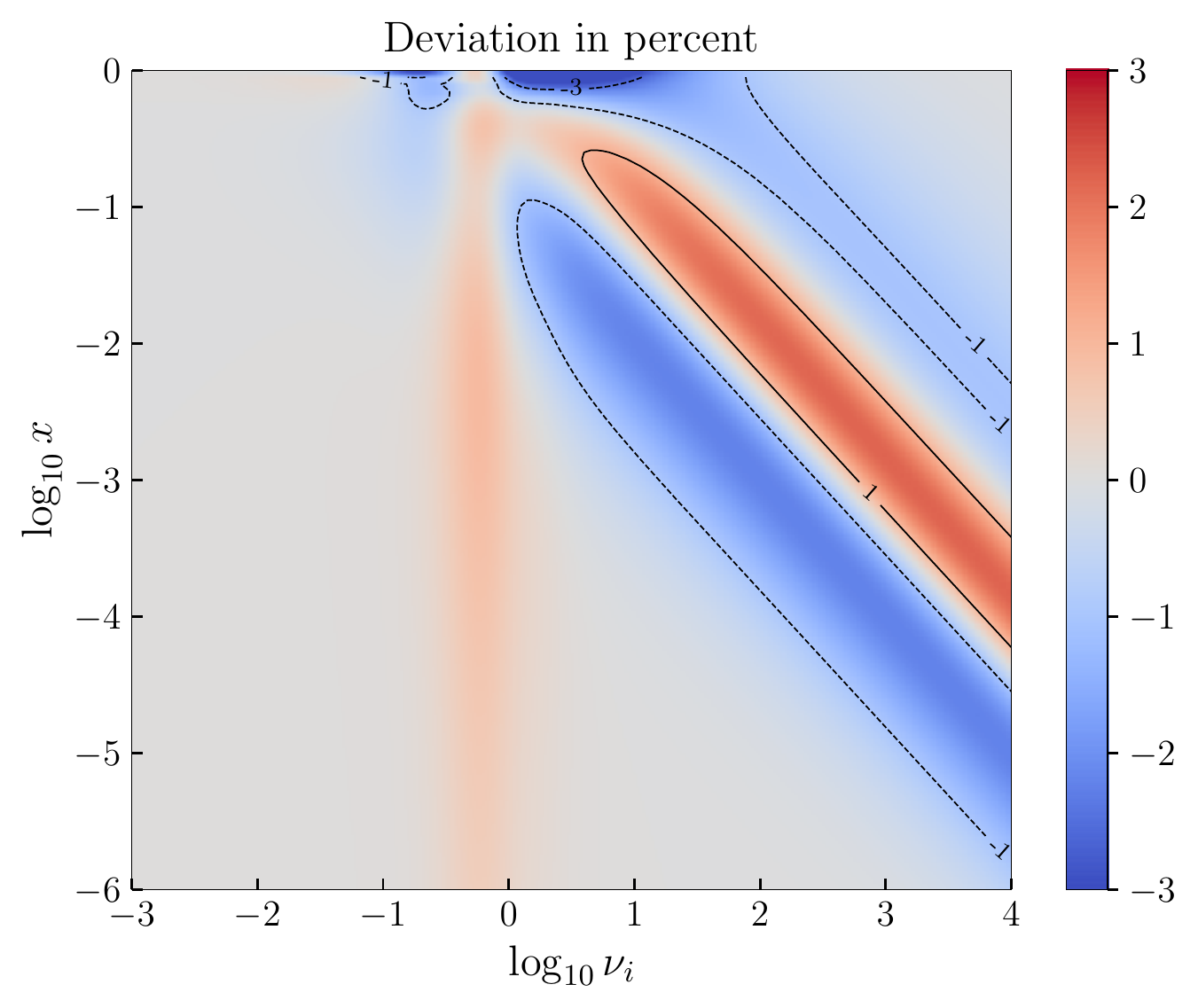}
		\caption{
		\textit{Left:} Regions of validity of several approximations  with better than 1\% error in the $\nu_i$, $x$ plane. The Elwert approximation~\eqref{eq:elwert} is shown in red, the soft-photon approximation~\eqref{eqn:weinberg} in gray and the Kramers limit ($\gff=1$) and the Kramers expansion~\eqref{eq:kramers_exp} in blue. 
		\textit{Right:} Relative error of the approximate formula~\eqref{eqn:approximate}. Regions where the approximation overestimates (underestimates) the Gaunt factor are shown in red (blue). The contours of positive (negative) deviations are shown in solid (dotted) black lines as labeled.
		\label{fig:regions}}
\end{figure*}

\subsection{Definition and previous results}

For electron-ion bremsstrahlung, the differential cross section is conventionally expressed as a product of the classical electrodynamics result by \cite{Kramers:1923} and a free-free Gaunt factor $g_{\rm ff}$ that corrects for the quantum nature of the process,
\begin{align}
\label{kramers}
    x \frac{d \sigma}{dx} &=
    \frac{16\pi}{3\sqrt{3}} \frac{\alpha^{3} Z^2 }{m_e^2 v_i^2} \times \gff (x,\nu_i).
\end{align}
The Gaunt factor is a function of the emitted photon energy $\omega$ measured in units of the available initial  energy $x$.%
\footnote{Treating the relative velocities $v_{i,f}$ as relativistic quantities (momentum over energy)  extends the validity of the non-relativistic expressions into the mildly relativistic region; see~\cite{Chluba:2019ser} for a quantitative comparison.} 

The exact non-relativistic Gaunt factor
 is most often quoted in the following form~\citep{1961ApJS....6..167K},%
 \footnote{In its shortest version, the Gaunt factor is written in terms of a derivative of the absolute square of a hypergeometric function~\citep{Sommerfeld:1935ab}, which is, however, less amenable to immediate numerical evaluation.}
\begin{subequations}
\label{eqn:full_quantum_cs}
\begin{align}
\gff = & \frac{2 \sqrt{3}}{\pi \nu_{i} \nu_{f}}\left[(\nu_{i}^{2}+\nu_{f}^{2}+2 \nu_{i}^{2} \nu_{f}^{2}) I_{0} \right. \nonumber \\ & \left. \quad -2 \nu_{i} \nu_{f}(1+\nu_{i}^{2})^{1 / 2}(1+\nu_{f}^{2})^{1 / 2} I_{1}\right] I_{0}
\end{align}
with the definitions of $I_0$ and $I_1$ given through
\begin{align}
I_{l} & = \frac{1}{4}\left[\frac{4 \nu_{i} \nu_{f}}{\left(\nu_{i}-\nu_{f}\right)^{2}}\right]^{l+1} e^{\pi|\nu_{i}-\nu_{f}| / 2} 
\nonumber \\
&\qquad \times \frac{|\Gamma\left(l+1+i \nu_{i}\right) \Gamma\left(l+1+i \nu_{f}\right)|}{\Gamma(2 l+2)} 
G_{l},
\end{align}
and where $G_{l}$ is the real function,
\begin{align}
G_{l}=|\beta|^{-i \nu_{i}-i\nu_f}{ }_{2} F_{1}(l+1-i \nu_{f}, l+1-i \nu_{i} ; 2 l+2 ;z) 
\end{align}
\end{subequations}
with $\beta = (\nu_i+\nu_f)/(\nu_i-\nu_f)$ and $z = -{4 \nu_{i} \nu_{f}}/{(\nu_{i}-\nu_{f})^{2}} $.

Equation~\eqref{eqn:full_quantum_cs} is exact in the Coulomb interaction and is therefore valid for arbitrary $\nu_i$ and~$\nu_f$.
In the classical regime, $\nu_{i,f} \gg 1$, or for hard photon emission $x\to 1$ for which $\nu_f \gg 1$, Eq.~\eqref{eqn:full_quantum_cs} requires the evaluation of ${}_2F_1$ for large imaginary arguments. This is a difficult and computationally expensive task~\citep{2008CoPhC.178..535M,2016arXiv160606977J} and in practical applications  either tabulations or expansions with limited range of validity are used.

Before we proceed with introducing a new and simple approximate formula, we collect the most important limits of and approximations to $\gff$ in Eq.~\eqref{eqn:full_quantum_cs}. 
 In the Born limit, $\nu_{i,f} \ll 1$, the Gaunt factor reduces to~\citep{Oppenheimer1929, Sugiura1929}
\begin{align} \label{eq:born_soft}
 \left. \gff \right|_{\text{Born}}
&= \frac{\sqrt{3}}{\pi} 
 \ln \left(\frac{1+ \sqrt{1-x}}{1- \sqrt{1-x}}\right) .
\end{align}
In the classical limit, $\nu_i \gg 1$, the asymptotic behaviour of the hypergeometric function depends on the product of $x$ and $\nu_i$, yielding two distinct cases. For $x\nu_i \ll 1$, Eq.~\eqref{eqn:full_quantum_cs} reduces to~\citep{landau1975classical}
\begin{equation} \label{eq:class_soft}
	\left.	\gff \right|_{\text{classical}}^{\text{soft}} =
		\frac{\sqrt{3}}{\pi}
		\ln\left(\frac{4}{x \, \nu_i \, e^{\gamma}}\right),
\end{equation}
where $\gamma = 0.5772\dots$ is the Euler-Mascheroni constant.
For $x\nu_i \gg 1$, one retrieves the result by \cite{Kramers:1923} and the Gaunt factor is  unity by definition.

Finally, there is a third limiting case, when the colliding electron is fast, $\nu_i\ll 1$, but when a significant fraction of energy is radiated for which $\nu_f\gg 1$,~\citep{berestetskii1982quantum}
\begin{equation} \label{eq:hard}
	\left.	\gff \right|_{\text{hard}} =   \frac{4\sqrt{3}\nu_i}{1-e^{-2\pi \nu_f}}.
\end{equation}
Similarly to the Kramers limit, this Gaunt factor tends to a finite value for $x\to 1$. This is owed to the possibility of bound-state formation by emission of quanta with $x>1$. Equation~\eqref{eq:hard} has a rather limited range of validity in the $(\nu_i,x)$ parameter space.

The two most important attempts to extend the regions of validity of the Born and classical expressions are the following:
in the prescription by~\cite{Elwert:1939km} one multiplies the Born expression by a ratio of Sommerfeld factors, 
\begin{align} \label{eq:elwert}
    \left.\gff\right|_\text{Elwert} &=
    \frac{\sqrt{3}}{\pi} \frac{\nu_f}{\nu_i}\frac{1-e^{-2\pi\nu_i}}{1-e^{-2\pi\nu_f}}
    \ln \left(\frac{1+ \sqrt{1-x}}{1- \sqrt{1-x}}\right) ,
\end{align}
which extends the region of validity of the Born result to $\nu_i\lesssim 1$.
In turn, the original classical expression by Kramers can be extended by expanding Eq.~\eqref{eqn:full_quantum_cs} for large values of $\nu_f-\nu_i$~\citep{10.1093/mnras/118.3.241},
\begin{align} \label{eq:kramers_exp}
    \left.\gff\right|_\text{Kramers exp.} &=
    1+\frac{0.21775}{\left(\nu_f-\nu_i\right)^{2/3}} - \frac{0.01312}{\left(\nu_f-\nu_i\right)^{4/3}}  .
\end{align}

Finally, in an insightful analysis (and clearing some previous misconceptions about widely used approximations to the Gaunt factor), \cite{Weinberg:2019mai} recently found a dipole formula for soft-photon emission, $x\ll 1$, that is valid for arbitrary~$\nu_i$.
    \begin{align} \label{eqn:weinberg}
    \left.\gff\right|_{\text{soft}} &=
    \frac{\sqrt{3}}{\pi} 
    \left[
    \ln \left(\frac{4}{x \zeta}\right) 
    + \frac{\pi^2 \nu_i^2}{\sinh^2{\pi \nu_i}} \ln \zeta
    \right].
\end{align}
Here, $\zeta = \nu_i e^{\gamma}$ is determined by matching onto the classical limit~\eqref{eq:class_soft}. 
The expression is obtained from an old formula that applies to soft-photon emission~\citep{Weinberg:1965nx} and the sum of two terms appears in a splitting of the electron scattering-angle integral; the factor $S_iS_f|_{x\to 0} = \pi^2\nu_i^2/\sinh^2{\pi \nu_i}$ is a correction introduced to the forward-scattering direction;
here, $S_{i,f} = \pm 2\pi \nu_{i,f} /(e^{ \pm 2\pi\nu_{i,f}}-1 )$ are the Sommerfeld factors where the positive and negative signs correspond to~$i$ and~$f$, respectively.
The Gaunt factor $\left.\gff\right|_{\text{soft}}$ asymptotes to $\left.	\gff \right|_{\text{classical}}^{\text{soft}}$ in the limit $\nu_i\gg 1$ and to $\left.	\gff \right|_{\text{Born}}^{\text{soft}} = \sqrt{3}/\pi 
 \ln \left(4/x\right)$ in the limit $\nu_i\ll 1$. The region of validity of Eq.~\eqref{eqn:weinberg} is  $x\ll \min\{1,\nu_i^{-1}\}$, limiting its applicability to soft photons, especially in the classical regime when~$\nu_i\gg 1$.

 In the left panel of Fig.~\ref{fig:regions} we summarize the most important approximations introduced above and show their regions of validity to within 1\% accuracy in the  $\nu_i$-$x$ plane.  One can see that the Elwert approximation~\eqref{eq:elwert} stays within 1\% accuracy for $\nu_i\lesssim 0.2$ and extends to larger values of $\nu_i$ towards the kinematic endpoint $x\to 1$. The classical expression by Kramers  ($\gff = 1$) becomes valid for $x \nu_i \gtrsim 160$ and the further expansion~\eqref{eq:kramers_exp} extends this region to $x \nu_i \gtrsim 0.6$. Finally, the soft-photon approximation~\eqref{eqn:weinberg}  stays within 1\%  accuracy for $x \nu_i \lesssim 0.007$ in the semi-classical regime and for $x\lesssim 0.08$ in the Born regime.

\subsection{A new approximate formula}

The soft-photon formula~\eqref{eqn:weinberg} is remarkable in that it is able to cover arbitrary~$\nu_i$, but, unfortunately, with a restriction on the smallness of~$x$, especially in the classical region. However, inspection of the  Born limit~\eqref{eq:born_soft} and of the Kramers classical expression~\eqref{kramers} suggests an extension that lifts those restrictions, widens the validity to $x\sim 1$ and includes the classical regime where $x \nu_i \gg 1$. We propose the formula,%
\begin{widetext}
\begin{align} \label{eqn:approximate}
       \gff &\simeq
    \frac{\sqrt{3}}{\pi} 
    \left[
    \ln \left(1+\frac{1}{\nu_i e^\gamma}\frac{1+\sqrt{1-x}}{1-\sqrt{1-x}}\right) 
    + \frac{\pi^2 \nu_i^2}{\sinh^2{\pi \nu_i}} \ln\left(\nu_i e^\gamma\right)
    - \frac{10}{3} \nu_i^2 \, e^{-\frac{3}{2}\pi \nu_i} \left(\frac{5}{1+e^{10(x-6/7)}}-2\right)
    \right]
    + \frac{x \nu_i }{\pi^{-2/3}+x\nu_i } . 
\end{align}
\end{widetext}
This equation yields the correct Born limit \eqref{eq:born_soft} for $\nu_i\ll 1$ for which the logarithms combine and all other contributions become negligible. For $\nu_i\gg 1$ it also approaches the correct classical limits \eqref{eq:class_soft} for $x\nu_i\ll 1$ and unity for $x\nu_i\gtrsim 1$. The additional term $+1$ in the first logarithm prevents the classical soft-photon cross section from yielding unphysical negative values for $x \nu_i \gtrsim 1$. In the same limit, the last term in Eq.~\eqref{eqn:approximate} ensures the matching onto the classical hard-photon  cross section for $x\nu_i\gg 1$. 
Finally, the third term in the square brackets is added by hand to improve on the quantum-to-classical crossover region $\nu_i\sim 1$ for which the soft-photon expression~\eqref{eqn:weinberg} deviates most from the exact result, even for $x\ll 1$. There, the term in the round brackets is approximately $+3$  for $x\lesssim 0.5$ and 
steeply drops to approximately
$-1$ towards the kinematic endpoint, changing sign at $x\simeq 0.9$.

It is clear that~\eqref{eqn:approximate} is not unique. For example, one may attempt to combine the benefits of the Elwert- and soft-photon approximation for a more precise coverage of the hard-photon region.
It turns out, that this generally requires stitching together terms using step-like functions and one tends to end up with unwieldy expressions. Our guiding principle is to provide a short and easy-to-evaluate analytical formula, even if it comes with small deductions on accuracy. As will be seen below, an additional benefit of~\eqref{eqn:approximate} is that it is readily integrable in~$x$.  In any case, if one strives for precision, one should of course use the exact result.

In the right panel of Fig.~\ref{fig:regions}, we show the relative error of Eq.~\eqref{eqn:approximate} with respect to the exact Gaunt factor~\eqref{eqn:full_quantum_cs} as a function of $x$ and $\nu_i$. Solid (dashed) contours delineate the regions where the approximation overestimates (underestimates) the exact result by more than 1\%. The diagonal positive and negative error regions in the left panel stem from the transition between the soft ($x\nu_i\ll 1$) and hard ($x\nu_i\gg 1$) classical limits. As can be seen, the approximation works generally better than 3\% in almost the entire parameter space. In the two isolated regions left and right of $\log_{10}\nu_i =0$ and close to the kinematic endpoint $x\to 1$, the deviations become largest. There, the relative error climbs above 3\% for $x>0.7$ and reaches its maximum of 17\% (10\%) at the very endpoint, $x = 0.99$, in the left (right) island. 

\begin{figure*}[tb]
		\includegraphics[width=\textwidth]{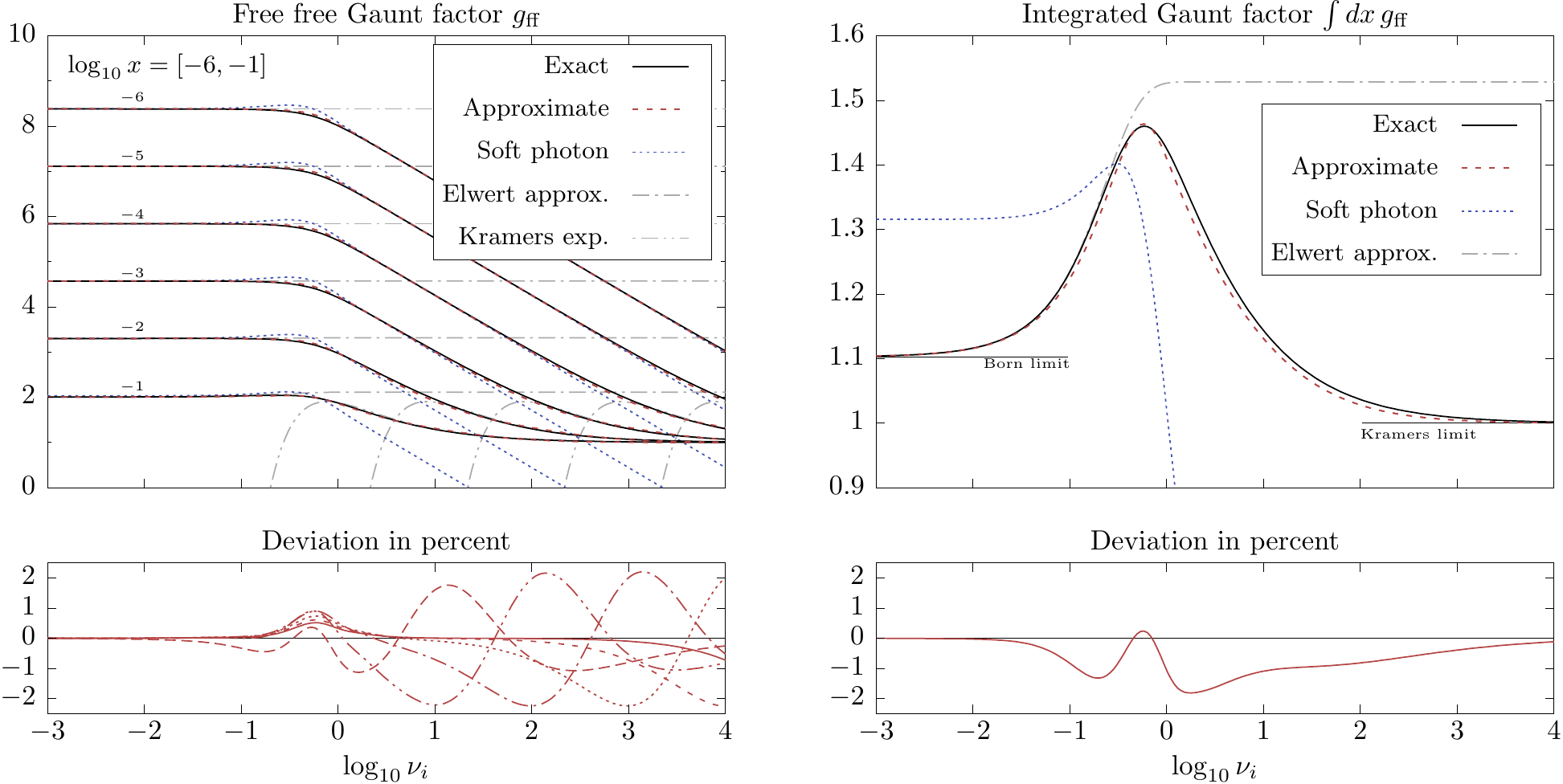}
		\caption{Free-free Gaunt factor (left) and integrated Gaunt factor (right) over a wide range of parameters $x$ and $\nu_i$. The approximate formula~\eqref{eqn:approximate} (dashed red) is compared to the exact result (solid black lines). The soft-photon approximation~\eqref{eqn:weinberg} is shown in dotted blue, the Elwert approximation~\eqref{eq:elwert} and the Kramers expansion~\eqref{eq:kramers_exp} are shown in gray. The relative error of the approximation is shown in the bottom panels. In the left panel, the \{solid, dashed, dotted, dash-dotted, dash-double-dotted, long-dashed\} lines correspond to values of $\log_{10} x = \{-6,-5,-4,-3,-2,-1\}$.  \label{fig:g_ff}}
\end{figure*}

In the top-left panel of Fig.~\ref{fig:g_ff}, we show the numerical comparison between Eq.~\eqref{eqn:approximate} (dashed red lines) and the exact Gaunt factor (solid black lines) on the same $\nu_i$-interval for six values of $\log_{10}x=[-6,-1]$ in increments of 1~dex. To demonstrate the failure of the Elwert approximation, the Kramers expansion and the soft-photon approximation outside their respective regions of validity, we also show the values of Eq.~\eqref{eq:elwert},~\eqref{eq:kramers_exp}  and~\eqref{eqn:weinberg} and in dash-dotted, dash-double-dotted and dotted lines respectively. 
In the bottom-left panel, the relative error of Eq.~\eqref{eqn:approximate} with respect to the exact Gaunt factor is shown. As can be seen, our approximation has a maximum error of 2.3\% for the chosen $x$-range.

\section{Thermally averaged Gaunt factors}
\label{sec:therm_gaunt}

\begin{figure*}[tb]
		\includegraphics[width=\textwidth]{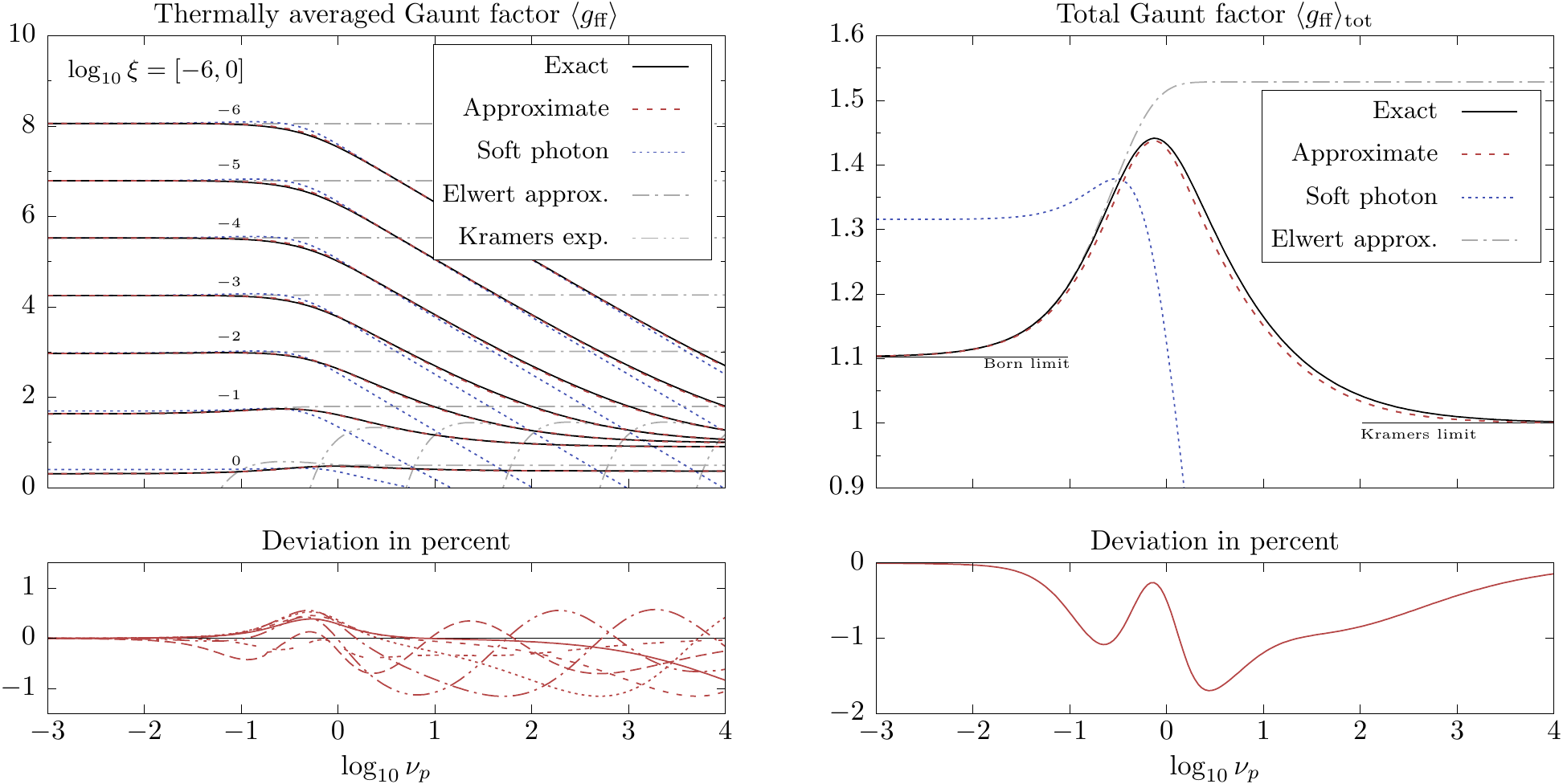}
		\caption{  
		Thermally averaged free-free Gaunt factor (left) and total Gaunt factor (right) over a wide range of parameters $\xi$ and $\nu_p$. The approximate formula~\eqref{eqn:approximate} (dashed red) is compared to the exact result (solid black lines). The soft-photon approximation~\eqref{eqn:weinberg} is shown in dotted blue, the Elwert approximation~\eqref{eq:elwert} and the Kramers expansion~\eqref{eq:kramers_exp} are shown in gray. The relative error of the approximation is shown in the bottom panels.
		In the left panel, the \{solid, dashed, dotted, dash-dotted, dash-double-dotted, long-dashed, triple-dotted\} lines correspond to values of $\log_{10} \xi = \{-6,-5,-4,-3,-2,-1,0\}$.
		\label{fig:g_ff_th}}
\end{figure*}

A non-relativistic Maxwellian plasma of  temperature $T$ is a commonly encountered environment in astrophysics. 
The average production of photons per volume, time and photon energy due to dipole bremsstrahlung  in such a plasma is given by
\begin{align}
  \frac{d\Gamma_{\rm brem}}{dV d\omega} &=  \frac{16}{3} \sqrt{\frac{2\pi}{3}} \frac{n_{e} n_{I} \alpha^{3} Z^2}{ m_e^{3/2} T^{1/2} \omega}   
   \langle\gff\rangle    .   
\end{align}
Here, $ \langle\gff\rangle  $ is a thermally averaged Gaunt factor and $n_e$ and $n_I$ are the number densities of electrons and (fully ionized) ions. Introducing
the dimensionless energy $\xi\equiv \omega/T$, the most probable value of $\nu_i$, $\nu_p \equiv  \alpha Z \sqrt{m_e/(2T)}$ and the dimensionless 
initial energy, $u\equiv m_e v_i^2/(2T)$, the averaged Gaunt factor is given by~\citep{1961ApJS....6..167K},
\begin{equation}
    \langle\gff\rangle (\xi, \nu_p)  =  \int_\xi^\infty du \, 
   e^{-u}\;  \gff\left(x = \frac{\xi}{u}  , \nu_i = \frac{\nu_p}{\sqrt{u}}\right),
\end{equation}

Another quantity of ample astrophysical interest is the cooling function, \ie, the energy lost per volume and time in the plasma,
\begin{align}
    \Lambda_\text{brem} = \frac{16}{3} \sqrt{\frac{2\pi}{3}} \frac{n_{e} n_{I} \alpha^{3} Z^2}{ m_e^{3/2} } T^{1/2}  \langle\gff\rangle_\text{tot}.
\end{align}
Here, $\langle\gff\rangle_\text{tot} $ is 
the total, energy-integrated and thermally averaged, Gaunt factor~\citep{1961ApJS....6..167K}
\begin{subequations}
\begin{align} \label{eq:def_gtot}
    \langle  \gff  \rangle_\text{tot} (\nu_p) &= 
    \int_0^\infty d\xi \,\langle  \gff  \rangle (\xi, \nu_p) \\
    \label{gtot_vtwo}
    & = \int_0^\infty du\, u\, e^{-u} \int_0^1  dx\, \gff\left(x,\nu_i=\frac{\nu_p}{\sqrt{u}} \right) .
\end{align}
\end{subequations}
The second line is particularly convenient because the $dx$ integration of Eq.~\eqref{eqn:approximate} can be performed analytically. One then  obtains $  \Lambda_\text{brem} $ by the single remaining $du$ integration of~\eqref{eqn:approximate_int} in~\eqref{gtot_vtwo}. The $dx$ integral yields 
\begin{widetext}
{
\medmuskip=0mu
\thinmuskip=1mu
\thickmuskip=1mu
\nulldelimiterspace=1pt
\scriptspace=1pt
\begin{align} \label{eqn:approximate_int}
    \int_0^1  dx\;\gff & =
    \frac{\sqrt{3}}{\pi} 
    \Bigg[  
    \frac{
    (1 + \nu_i e^\gamma )^2\ln\left(1+\nu_i e^\gamma\right)+
    2\left(1-\nu_i e^\gamma\left(1+\ln 4\right)\right)
    }{(1-\nu_i e^\gamma)^2} 
    - \ln\left(\nu_i e^\gamma\right)
    + \frac{\pi^2 \nu_i^2 \ln \left(\nu_i e^\gamma\right)}{\sinh^2{\pi \nu_i}} 
    - A \, \nu_i^2 \, e^{-\frac{3}{2}\pi \nu_i}
    \Bigg]
   -  \frac{\ln\left(1+\pi^{2/3} \nu_i\right)}{\pi^{2/3} \nu_i}  +1 
    ,
\end{align}
}%
\end{widetext}
with $A=-5/21(18+7\log[(1+e^{10/7})/(1+e^{60/7})]) \simeq 7.26$. 
 For completeness, we also record the Born and classical limits, $ \int_0^1 \,  dx\, \left. \gff \right|_{\text{Born}} = {2 \sqrt{3}}/{\pi}$ and 
 $ \int_0^1 \, dx\, \left. \gff \right|_{\text{classical}} = 1  $, respectively.

In the top-right panel of Fig.~\ref{fig:g_ff}, we show the integrated approximate Gaunt factor~\eqref{eqn:approximate_int} and compare it to the numerical values of the exact expression.
In the bottom panel, the relative error of the approximation with respect to the exact Gaunt factor is shown. Equation~\eqref{eqn:approximate_int} has a maximum deviation of 1.8\% from the exact result.
The Elwert approximation~\eqref{eq:elwert} works well for $x\ll 1$ but overestimates the integrated Gaunt factor by more than 150\% in the classical regime.
The Kramers expansion%
\footnote{
The integrated Kramers expansion takes on negative values across the whole $\nu_i$ range in the right panel and is therefore not shown in Fig.~\ref{fig:g_ff}; the same holds for Fig.~\ref{fig:g_ff_th}.
}%
~\eqref{eq:kramers_exp} and the soft-photon approximation~\eqref{eqn:weinberg} fail completely when the $x$-integration is performed since they take on negative values outside their region of validity.

In Fig.~\ref{fig:g_ff_th} we show the thermal average of Eq.~\eqref{eqn:approximate} and Eq.~\eqref{eqn:approximate_int} and compare it to the exact result.
In the top-left panel, the approximate thermally averaged Gaunt factor $\langle\gff\rangle$ is shown in dashed red and the exact one in solid black for seven values of $\log_{10}\xi=[-6,0]$ in increments of 1~dex.
The Elwert approximation~\eqref{eq:elwert}, the Kramers expansion~\eqref{eq:kramers_exp} and the soft-photon approximation~\eqref{eqn:weinberg} are given by the dash-dotted, dash-double-dotted and dotted lines, respectively.
In the top-right panel, the same comparison is shown for the total Gaunt factor $\langle\gff\rangle_\text{tot}$. 
In the bottom panels one observes that the maximum error of the approximate Gaunt factor is 1.1\% for $\langle\gff\rangle$. The approximate total Gaunt factor $\langle\gff\rangle_\text{tot}$ is always smaller than the exact result and has a maximum error of 1.7\%.

\section{Conclusions}
\label{sec:conclusions}

In this work, we present an approximate formula for the Gaunt factor of electron-ion bremsstrahlung that is easy to evaluate and deviates less than 3\% from the exact result for most of the allowed parameter space. Larger deviations are only observed in isolated regions at the quantum-to-classical crossover when the photon energy is close to the kinematic endpoint. 
Our approximate formula is analytically integrable in photon energy. The integrated Gaunt factor has a maximum error of 1.8\% at $\nu_i\sim 1$. Both the approximate formula and the integrated version show the correct asymptotic behaviors in the Born and classical regions.

The two analytical approximate formul\ae\ allow for easy numerical integration to calculate thermally averaged quantities, such as production spectra and cooling rates. While production spectra can be estimated with 1\% accuracy, the approximate bremsstrahlung-induced cooling rate is smaller than or equal to the exact result across the entire phase space with a maximum error below 2\% reaching its maximum at the quantum-to-classical crossover. While our proposed expressions cannot replace exact results (or highly-accurate tabulations thereof), they are helpful for numerically inexpensive yet reasonably accurate determinations of bremsstrahlung rates of arbitrary non-relativistic relative velocity and all photon energies.

\paragraph*{Acknowledgements: }\ \  The authors are supported by the New Frontiers program of the Austrian Academy of Sciences. LS was in part supported by the Austrian Science Fund FWF under the Doctoral Program W1252-N27 Particles and Interactions.

\bibliography{References.bib}
\end{document}